\documentclass{jpsj2}
\def\alt{\lesssim}
\def\agt{\gtrsim}

\title{
A Langevin Approach to One-Dimensional Granular Media Fluidized by Vibrations
}

\author{
Jun'ichi \textsc{Wakou}$^{1}$\thanks{E-mail address: wakou@cc.miyaknojo-nct.ac.jp},
Akinori \textsc{Ochiai}$^{2}$ and
Masaharu \textsc{Isobe}$^{2}$\thanks{E-mail address: isobe@nitech.ac.jp}
}

\inst{
$^{1}$Miyakonojo National College of Technology, Miyakonojo-shi, Miyazaki, 885-8567 \\
$^{2}$Graduate School of Engineering, Nagoya Institute of Technology, Nagoya 466-8555
}

\abst{
We present a Langevin approach to describe the steady-state dynamics of one-dimensional granular media fluidized by a vibrating bottom plate.
We adopt a linear Langevin equation to describe the motion of the center of mass. Within this framework, we derive analytical expressions for several macroscopic quantities. 
We also predict the power spectrum for the height of the center of mass.
We find good agreement between our theoretical predictions and extensive event-driven molecular dynamics simulations.
}

\kword{granular matter, vertical oscillation, fluidization, scaling relation, 
center of mass, Langevin equation, event-driven molecular dynamics simulation
}

\begin{document}
\maketitle

\section{Introduction}
Granular materials fluidized by external vibrations have attracted a great deal of interest in the field of nonequilibrium statistical physics.
These systems show a variety of fascinating phenomena (see, for example, ref.~\citen{aranson_2006}), such as convection, surface waves, and size segregation.
The transition from a fluidized state to a condensed state, in which the particles move collectively with the same period as the external vibrations, has been studied both experimentally~\cite{clement93,luding94,goldshtein95} and numerically~\cite{clement93,luding94,esipov99}.

Scaling laws of the macroscopic properties ({\em e.g.} the center of mass (COM)) in a fluidized state are important subjects from the viewpoint of a nonequilibrium steady state (NESS), and have also been extensively studied using experiments~\cite{luding94,warr95,wildman01}, simulations~\cite{luding94,luding94-2,luding95,mcnamara98}, kinetic theory~\cite{bernu94,warr95,kumaran98,kumaran98-2}, and hydrodynamic descriptions~\cite{lee95,lee97}.
However, no agreement has been reached among the various studies for the scaling relationship of the height of the COM in a NESS. 
This discrepancy is not yet understood and the question remains unanswered.

To resolve this problem, we focused on a one-dimensional column of inelastic hard rod (granular) particles bouncing on a sinusoidally vibrating bottom plate.
A one-dimensional system allows us to study the pure dynamics of the COM in a fluidized state since we can remove the effect of convection within granular beds, which becomes dominant in higher dimensions.
A one-dimensional system has been investigated in detail previously by several authors.
Numerical simulations~\cite{luding94} have clarified the behavior of the transition from a condensed state to a fluidized state and showed a scaling relationship for the height of the COM in the fluidized state.
The transition is characterized by a parameter $X\equiv N(1-r)$, where $N$ is the number
of granular particles and $r$ is the restitution coefficient for collisions between
particles. If $X$ is greater than the critical value $X_C$, which is about 3, then
the transition does not occur regardless acceleration of the vibrating bottom plate;
if $X<X_C$, then the transition may occur at a certain magnitude of the acceleration.
As an explanation of the value of $X_C\simeq 3$, only connection with the result
of an analytical study~\cite{bernu90} on inelastic particles colliding with a wall
has been pointed out. 
The scaling relationship for the height of the COM has been shown to be characterized
by $X$ and typical velocity of the bottom plate.
Theoretical considerations~\cite{bernu94} based on kinetic theory have shown that analytical solutions of a dissipative Boltzmann equation for a one-dimensional system agree well with the results of numerical simulations.

The main purpose of our study was to develop a novel theoretical method to clarify the macroscopic behavior of fluidized granular media through the dynamics of the COM without using a kinetic equation.
Our theoretical method has a wide range of applications to both gas states, for which kinetic theory can be applied, and dense liquid states, when the COM of the granular media evolves with the same period of the bottom plate.
Our basic formulation, which consists of a Langevin equation of motion of the COM, was derived by focusing on the force acting on the granular media due to the bottom plate.
Some macroscopic quantities can be derived easily from the solution of this equation.
Although several phenomenological parameters must be estimated by numerical simulation, we confirmed that our compact formulation agreed with extensive event-driven molecular dynamics simulations self-consistently.

We considered a column of $N$ particles of mass $m$ bouncing on a vibrating bottom plate subjected to gravity with acceleration $g$.
The motion of the particles was confined to the $z$ axis, and the $z$ direction was opposite to the direction of gravity.
Since the diameter of the particles plays no role in one-dimensional hard sphere systems, we regarded the particles as point particles.
The bottom plate oscillated sinusoidally with amplitude $A_0$ and angular frequency $\omega$; hence, the height of the bottom plate $z_0(t)$ at time $t$ was \begin{eqnarray}
z_0(t)=A_0\sin\left(\omega t\right).
\end{eqnarray}
Collisions between particles were inelastic with a restitution coefficient $r$. 
For simplicity, we assumed that collisions between the lowest particle and the bottom plate were elastic.

Two important timescales occur in this system: the oscillation period of the bottom plate $\tau (\equiv 2\pi/\omega)$, and the macroscopic relaxation time $\tau_{rel}$ to the stationary state.
If $\tau$ is comparable to $\tau_{rel}$, the energy supplied by one stroke of the bottom plate is almost dissipated during the period, and the particles will be in a condensed state.
Such a condensed state is beyond the scope of the present study.
In this study, we restricted ourselves to the high-frequency case $\tau/\tau_{rel}\ll 1$, in which the system is in a fluidized state.

We performed simulations systematically by changing the number of particles $N$ ($N=10, 100, 1000$), the restitution coefficient $r$ ($r=0.80\sim 0.9999$), and the maximum acceleration of the bottom plate $\Gamma\equiv A_0\omega^2/g$ ($\Gamma=10\sim 640$).
The physical quantities were averaged over a long period of time: $1.0\times 10^5\tau$ for $N=10$ and $100$, and $5.0\times10^4\tau$ for $N=1000$.

\section{A Langevin Approach}
To describe the macroscopic properties of the fluidized state, we started from the equation of motion of the COM:
\begin{eqnarray}
M\frac{d^2 Z}{dt^2}=-Mg+F_b ,
\label{eqmotion}
\end{eqnarray}
where $Z(t)$ is the height of the COM of the particles at time $t$, $M=Nm$ is the total mass, and 
the right hand side of eq.~(\ref{eqmotion}) represents the sum of all 
external forces acting on
the granular particle system: the first term is the gravitational force
and the second term $F_b(t)$ is the external force
exerted by the vibrating bottom plate at time $t$.
The long-time average of $F_b(t)$ must balance with the gravitational force $Mg$ acting on the column of particles.
Hence, the central part of this study evaluates the deviation of $F_b(t)$ from its long-time average:
$\delta F(t)=F_b(t)-Mg$.

To evaluate $\delta F(t)$, we focused on the reaction force of $F_b(t)$, 
{\it i.e.}, the force exerted on the bottom plate by the granular fluid. 
We drew an analogy
between the force acting on the bottom plate and the force acting on a fine particle exhibiting Brownian motion while immersed in a fluid (see, for example, ref.~\citen{kubo85}); here the bottom plate on one side of the granular fluid
corresponds to a Brownian particle in a fluid~\cite{comment1}.
The force on one side of the Brownian particle consists of the pressure, 
the frictional force, and the random force.
We assumed that the force on the bottom plate also contained these three characteristic
forces: the average force $Mg$ on the bottom plate corresponding to the average pressure
on the Brownian particle, the frictional force accompanied by the relative motion
of the bottom plate against the granular fluid, and the random force. 
In addition we have to take into account that time dependence of the pressure
which is originated from the elasticity of the granular fluid. We assumed here
two kinds of elastic forces which cause time dependence of the pressure.
The first is the reaction force when the bottom plate excites a sound wave
in the granular fluid.
The second is the elastic force accompanied by macroscopic motion of the granular
fluid; this force is significant in our system because the granular fluid of 
finite length has the slowest oscillating mode with a macroscopic time scale.

Finally, we assumed the following form of $\delta F(t)$, consisting of systematic forces and a random force:
\begin{eqnarray}
\delta F(t)=-k\left(Z(t)-\overline{Z}\right)
+f_s(t)-\mu V(t) +R(t).
\label{deltaf}
\end{eqnarray}
The first term represents the elastic force accompanied by the slowest mode showing the expansion and contraction for the total length of the column of particles.
We assumed this was proportional to the deviation of the height of the COM $Z(t)$ from its stationary value $\overline{Z}$, $Z(t)-\overline{Z}$, where $\overline{Z}$ is defined by $\overline{Z}\equiv \lim_{T\to\infty}\frac{1}{T}\int_{0}^{T} Z(t)dt$.
The constant $k$ is the elastic constant.
The second term in eq.~(\ref{deltaf}) is the elastic force $f_s(t)$ accompanied by the excitation of a sound wave at the bottom plate.
The third term is the frictional force, which we assumed was proportional to the COM velocity $V(t)$.
The constant $\mu$ is the frictional constant. The last term $R(t)$ is the random force, which is specified later.

We estimated the elastic force $f_s(t)$ on the basis of hydrodynamic sound-wave theory~\cite{landau87}.
In a normal fluid, sound waves propagate according to a relationship between the pressure and the velocity of the fluid.
Let us denote a small change in the pressure from its equilibrium value by $p'$, a typical velocity of the fluid particles in the wave by $v$, and the velocity of sound by $c_s$.
If the condition $v\ll c_s$ is satisfied, we have a relationship $p'=\rho c_s v$ for a traveling plane wave, where $\rho$ is the constant equilibrium density of the fluid.
We assumed that this relationship was also satisfied in fluidized granular media under the same condition $v\ll c_s$.
Let us introduce here the global temperature $T$ and the thermal velocity $c$
of the granular fluid
defined in analogy with those of a normal fluid. The global temperature
is related to the mean square velocity fluctuation by 
$k_B T\equiv m\overline{\langle (v-\langle v \rangle_N)^2 \rangle_N}$,
where $k_B$ is the Boltzmann constant and the angle brackets 
$\langle\cdots \rangle_N$ denote the average over all particles. 
The thermal velocity $c$ is defined using the global temperature $T$ as 
$c=\sqrt{k_B T/m}$.
The velocity of sound $c_s$ is on the order of the thermal velocity $c$.
The density $\rho$ is on the order of $M/(c^2/g)$, where $c^2/g$ is the maximum height reached by a particle launched from the bottom plate with a thermal velocity $c$; this maximum height characterizes the length of the column of particles.
In the vicinity of the bottom plate, $v$ may be approximated by the velocity of the bottom plate $v_0(t)=A_0\omega\cos(\omega t)$.
Since $f_s(t)$ corresponds to $p'$ at the bottom plate, we have
\begin{eqnarray}
f_s(t)=\hat{\sigma}M\frac{g}{c}A_0\omega\cos(\omega t),
\label{fs}
\end{eqnarray}
where $\hat{\sigma}$ is a numerical factor on the order of 1 that is used as a curve-fit parameter when we compare our theoretical predictions with the results of simulations.

There are some important consequences derived from the condition $v\ll c_s$.
First, in the vicinity of the bottom plate, 
the condition $v\ll c_s$ can be written as $v_0\ll c$.
Hence, the maximum value of $v_0(t)$, $A_0\omega$, must be small compared to $c$: $A_0\omega\ll c$. Secondly, the condition $v\ll c_s$ suggests 
$\overline{\langle v \rangle_N^2}\ll c^2=\overline{\langle v^2 \rangle_N}-\overline{\langle v \rangle_N^2}
\simeq \overline{\langle v^2 \rangle_N}$.
Hence, 
\begin{eqnarray}
\frac{k_B T}{2}&=&\frac{m}{2} \left(\overline{
\langle v^2\rangle_N}-\overline{\langle v\rangle_N^2}
\right)
\simeq \frac{m}{2}\overline{\langle v^2\rangle_N}\equiv \overline{E}_K,
\end{eqnarray}
where $\overline{E}_K$ is the stationary value of the kinetic energy per particle
defined as 
$\overline{E}_K\equiv \overline{\frac{1}{N}\sum_{i=1}^{N}\frac{m}{2}v_i^2}$.
In our simulation, we estimated the thermal velocity $c$ according
to the relation $c=\sqrt{2\overline{E}_K/m}$.

The elastic constant $k$ is related to the angular frequency $\Omega$ of the slowest mode of the macroscopic oscillatory motion by $\Omega\equiv \sqrt{k/M}$. 
The period $\tau_{osc}$ of the oscillation is given by $\tau_{osc}\equiv 2\pi/\Omega$.
The rate of relaxation to the stationary state $\mu'$ due to friction is defined by $\mu'=\mu/M$, which is related to the macroscopic relaxation time $\tau_{rel}$ by $\tau_{rel}=\mu'^{-1}$.
Since both $\tau_{osc}$ and $\tau_{rel}$ characterize the macroscopic change that extends to the full length of the column of particles, they must be on the same order as the characteristic time taken for a sound wave to travel along the total length of the column of particles.
This characteristic time can be estimated as $c/g$ because the velocity of sound is on the order of $c$ and the total length of the column of particles is on the order of $c^2/g$.
Thus, we assume that $\Omega$ and $\mu'$ are on the order of $g/c$:
\begin{eqnarray}
\Omega=\hat{\Omega}\frac{g}{c},
\hspace{1cm}
\mu'=\hat{\mu}\frac{g}{c},
\label{omegaandmu}
\end{eqnarray}
where $\hat{\Omega}$ and $\hat{\mu}$ are numerical factors on the order of 1 that are determined by curve-fitting the results of our simulations.

We supposed that the properties of the random force $R(t)$ were the same as those acting on a fine particle undergoing Brownian motion in a fluid at temperature $T$~\cite{kubo85}.
Thus, $R(t)$ is stationary Gaussian white noise:
\begin{eqnarray}
\left<R(t)\right>=0,
\hspace*{1cm}
\left<R(t)R(t')\right>=I\delta(t-t').
\end{eqnarray}
We assumed that the fluctuation--dissipation theorem was satisfied:
\begin{eqnarray}
I=2M\mu'k_B T=2Mg\hat{\mu}m c,
\end{eqnarray}
where we used eq.~(\ref{omegaandmu}) and the relationship $c=\sqrt{k_BT/m}$ to derive the second equality.

Collecting the above results, we have the following linear Langevin equation for the COM:
\begin{eqnarray}
\frac{dV}{dt}=-\Omega^2\left(Z-\overline{Z}\right)-\mu' V +\frac{f_s}{M}+\frac{R}{M}.
\label{lang2}
\end{eqnarray}
This equation has the same form as the Langevin equation that describes the forced oscillations of a fine particle undergoing Brownian motion in a harmonic potential.
The solution to this equation has the form 
\begin{eqnarray}
Z(t)-\overline{Z}&=&A_0\zeta\sin(\omega t +\theta)
+\int_{-\infty}^{t}G(t-t')\frac{R(t')}{M}dt'
\nonumber\\
&&+F_{ini}(t),
\label{cmheight}
\end{eqnarray}
where
\begin{eqnarray}
\zeta&=&\frac{\hat{\sigma}\,\frac{g}{c}\,\omega}{\sqrt{(\Omega^2-\omega^2)^2+(\mu'\omega)^2}},
\end{eqnarray}
and
\begin{eqnarray}
\tan\theta &=& -\frac{\omega^2-\Omega^2}{\mu'\omega}
\hspace{0.5cm}
\left(-\frac{\pi}{2} \le \theta < 0 \right),
\end{eqnarray}
respectively.
The function $G(t)$ is given by
\begin{eqnarray}
G(t)&=&\frac{e^{-\frac{\mu'}{2}t}}{\omega_0}
\sin\left(\omega_0 t\right),
\label{gt}
\end{eqnarray}
where $\omega_0=(\Omega^2-(\mu'/2)^2)^{1/2}$.
The last term $F_{ini}(t)$ consists of those that depend on the initial conditions and vanish after a sufficient amount of time. 
Thus, the term is negligible when calculating long-time averages of physical quantities in a stationary state.

Let us consider a fluidized state with the timescale $\left(\tau/\tau_{rel}\right)^2\ll 1$.
This can be rewritten by substituting $\tau=2\pi/\omega$ and $\tau_{rel}=c/g$ as $\hat{\omega}^2\gg 1$, where $\hat{\omega}$ is defined as $\hat{\omega}=\omega c/g$.
In this limit, we expanded $\zeta$ in terms of $\hat{\omega}^{-2}$:
\begin{eqnarray}
\zeta
&=&\frac{\hat{\sigma}}{\hat{\omega}}
\left(1+O\left(\hat{\omega}^{-2}\right)\right).
\label{zeta2}
\end{eqnarray}
\begin{figure}[t]
\begin{center}
\includegraphics[height=5.0cm]{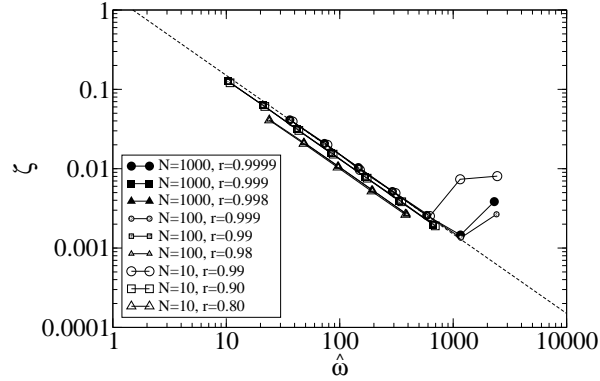}
\caption{\label{fig:zeta} A test of the theoretical results given by eq.~(\ref{zeta2}).
The accelerations $\Gamma$ used in the simulations were $\Gamma=10, 20, 40, 80, 160, 320$, and $640$, except for the simulations that experienced an "inelastic collapse" (an infinite number of collisions in a finite time)~\cite{bernu90,mcnamara92}.
The dashed line corresponds to the theoretical prediction $\hat{\sigma}/\hat{\omega}$
with $\hat{\sigma}=1.5$.}
\end{center}
\end{figure}
Figure~\ref{fig:zeta} gives the simulation results for $\zeta$ as a function of $\hat{\omega}$ with $c=(2\overline{E}_K/m)^{1/2}$.
As shown in previous studies~\cite{clement93,luding94,bernu94}, the relevant parameter that governs the behavior of the system was $X\equiv(N-1)(1-r)$
rather than $N$ or $r$; $X$ is the parameter that  
measures effective dissipation in the column of granular particles.
The factor $N-1$ in the definition of $X$ 
represents the number of dissipative contacts in the system, where $-1$ is due
to the fact that we assumed elastic collisions between the lowest particle
and the bottom plate.
Data points with the same $\Gamma$ and $X$ coincided in the figure.
For simulations with $X\alt 1$, shown by circles and squares, the master curve was consistent with the theoretical predictions (\ref{zeta2}) using $\hat{\sigma}=1.5$.
For $X\agt 1$, the numerical factor $\hat{\sigma}$ weakly depended on $X$.
We also observed deviations from the master curve at large $\hat{\omega}$ for the simulations with small $X$.
These deviations may be attributable an insufficient simulation time to obtain the stationary average.

\begin{figure}[t]
\begin{center}
\includegraphics[height=5.0cm]{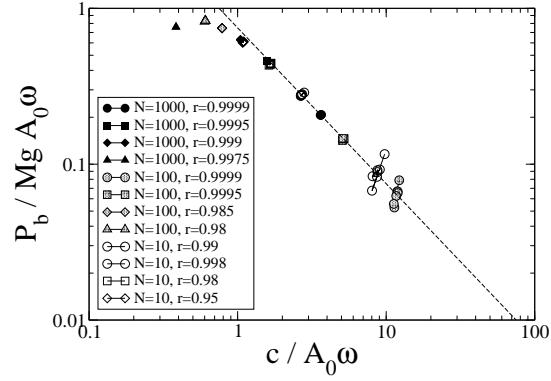}
\caption{\label{fig:scpb} A test of the theoretical result given by eq.~(\ref{pbsc}).
The accelerations $\Gamma$ used in the simulations were $\Gamma=10, 20, 40, 80, 160, 320$, and $640$, except the simulations that experienced an inelastic collapse.
The dashed line corresponds to the theoretical predictions $\left(\hat{\sigma}/2\right)\left(A_0\omega/c\right)$ with $\hat{\sigma}=1.5$.
}
\end{center}
\end{figure}

Some macroscopic quantities can be calculated using the formula given by eq.~(\ref{cmheight}).
Let us first consider the power injected by the bottom plate $P_b$, which has been the subject of recent studies~\cite{warr95,kumaran98,kumaran98-2,mcnamara97,soto04}.
Using the force supplied by the bottom plate $F_b$ and its velocity $v_0$, the power input $P_b$ can be defined by
$P_b=\overline{F_b v_0}=\lim_{T\to\infty}\frac{1}{T}\int_{0}^{T}F_b(t) v_0(t)dt$.
Substituting $F_b$ using eq.~(\ref{eqmotion}) in the definition of $P_b$ and integrating by parts twice, we have
\begin{eqnarray}
P_b 
&=& -M\omega^2
\lim_{T\to\infty}\frac{1}{T}\int_{0}^{T}
Z(t)v_0(t) dt.
\label{pb}
\end{eqnarray}
Then, substituting eq.~(\ref{cmheight}) into eq.~(\ref{pb}), we obtain
\begin{eqnarray}
P_b/MgA_0\omega&=&\frac{\hat{\sigma}}{2}\frac{A_0\omega}{c}
\frac{\omega^2\left(\omega^2-\Omega^2\right)}
{\left(\omega^2-\Omega^2\right)^2+\left(\mu'\omega\right)^2}
\nonumber\\
&=&\frac{\hat{\sigma}}{2}\frac{A_0\omega}{c}
\left(1+O\left(\hat{\omega}^{-2}\right)\right).
\label{pbsc}
\end{eqnarray}
The results obtained by neglecting terms on the order of $\hat{\omega}^{-2}$ coincided with the scaling predicted by kinetic theories~\cite{warr95,kumaran98,kumaran98-2}: $P_b\sim Mg\left(A_0\omega\right)^2/c$. Figure~\ref{fig:scpb} shows that the scaling relationship (\ref{pbsc}) with a factor $\hat{\sigma}=1.5$ agrees well with the simulations in the range when $A_0 \omega/c \alt 1$.
This result is consistent with the condition $A_0 \omega/c \ll 1$ required for the formula given by eq.~(\ref{fs}) to be valid.

In previous studies~\cite{warr95,kumaran98,kumaran98-2}
on the basis of kinetic theory, the scaling relationship for the granular temperature
$T\sim \left(A_0\omega\right)^2/X$
was derived from a balance between the power input from the bottom plate $P_b$
and the rate of energy dissipation due to inelastic collisions between particles.
If we assume the scaling relationship is satisfied in the parameter range
where our simulations were performed, we have $c\sim A_0\omega/\sqrt{X}$.
Then, the condition $A_0\omega/c\alt 1$ can be rewritten as $\sqrt{X}\alt 1$.
This result is consistent with the observation that the simulation data agreed well with the theoretical predictions in Fig.~\ref{fig:zeta} if $X\alt 1$.

Next, let us consider the power spectrum $I_{CM}$ for the height of the COM.
According to the Wiener--Khinchin theorem, this can be calculated analytically from the Fourier transform of the two-time correlation function $\psi_{CM}(t)$ defined by
\begin{eqnarray}
\psi_{CM}(t)=\lim_{T\to\infty}\frac{1}{T}\int_{0}^{T}
\left< \delta Z(t')\delta Z(t'+t)\right>dt'\,
,
\label{psicm}
\end{eqnarray}
where $\delta Z(t)\equiv Z(t)-\overline{Z}$ and the brackets $\left<\cdots\right>$ indicate an average over the random force $R(t)$.
Substituting eq.~(\ref{cmheight}) into eq.~(\ref{psicm}) and performing the Fourier transform gives
\begin{eqnarray}
I_{CM}(\hat{\omega'})/ \frac{c^5}{Ng^3}
&=&
\frac{\pi}{2}N\zeta^2
\left(\frac{Ag}{c^2}\right)^2
\left(\delta(\hat{\omega'}-\hat{\omega})
+\delta(\hat{\omega'}+\hat{\omega})\right)
\nonumber\\
&+&
\frac{2\hat{\mu}}{(\hat{\Omega}^2-\hat{\omega'}^2)^2+(\hat{\mu}\hat{\omega'})^2}
,
\label{pscmhsc}
\end{eqnarray}

\begin{figure}[H]
\begin{center}
\includegraphics[height=6.cm,clip]{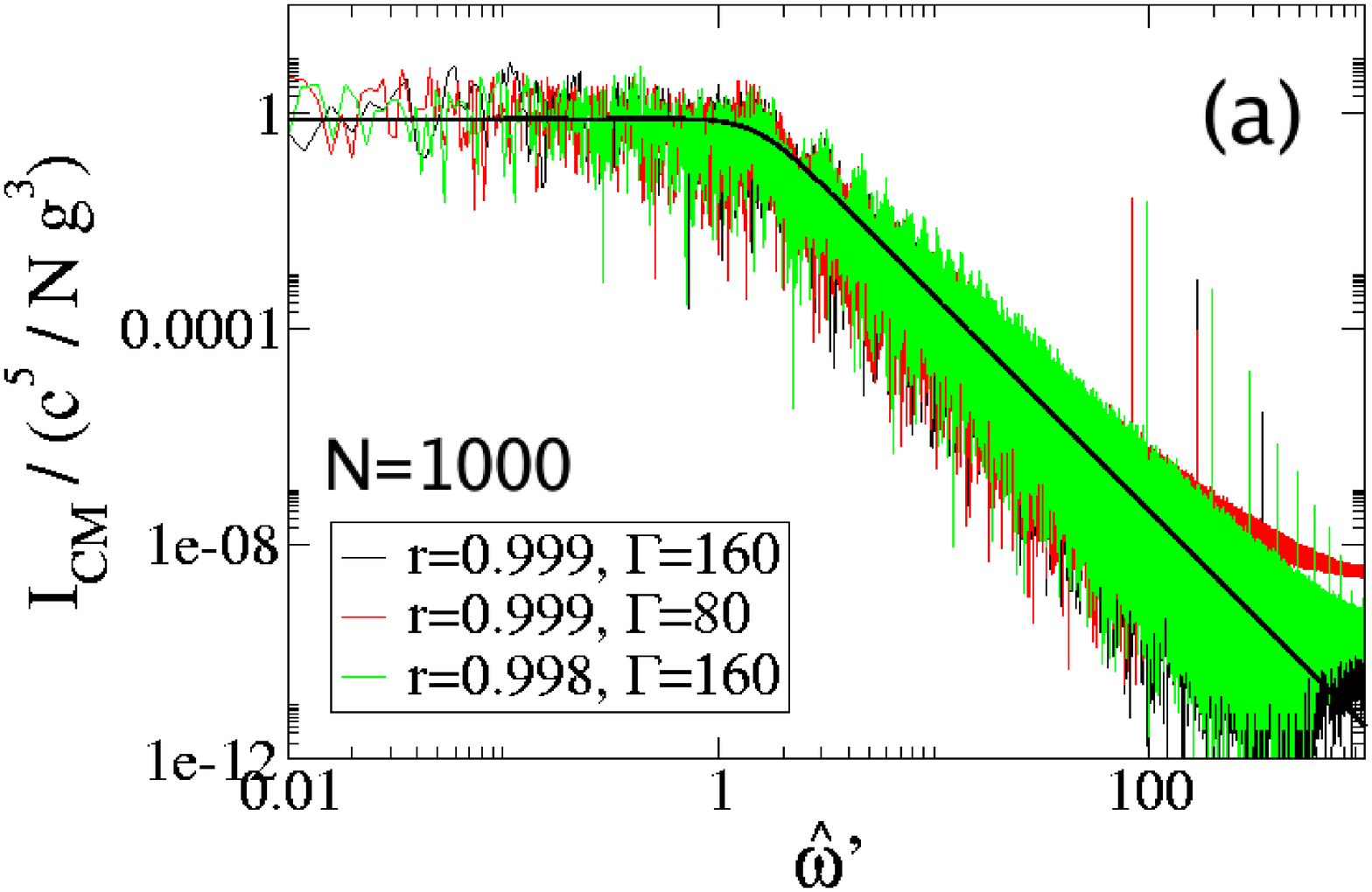}
\\[0.7cm]
\includegraphics[height=6.cm,clip]{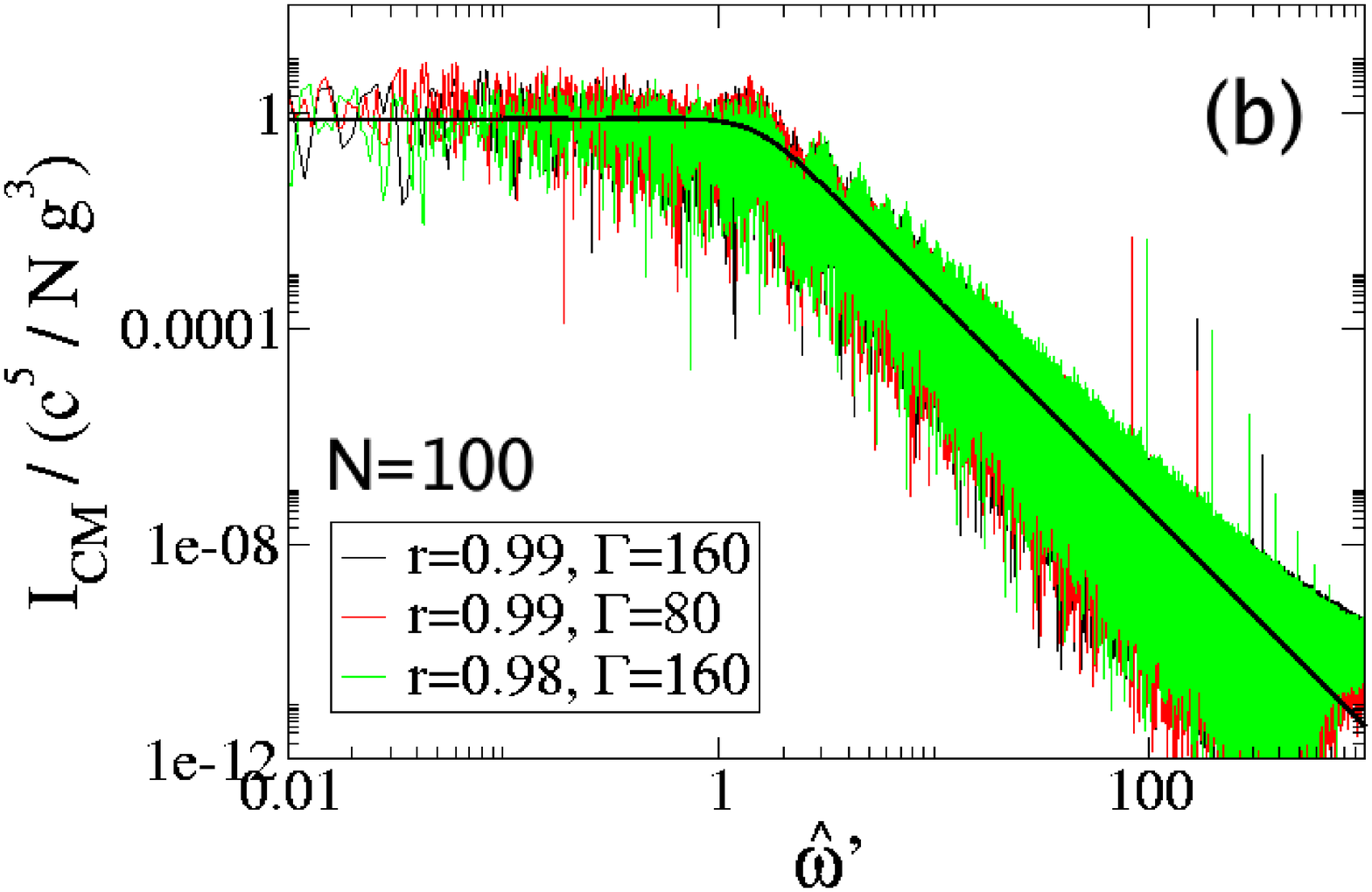}
\\[0.7cm]
\includegraphics[height=6.cm,clip]{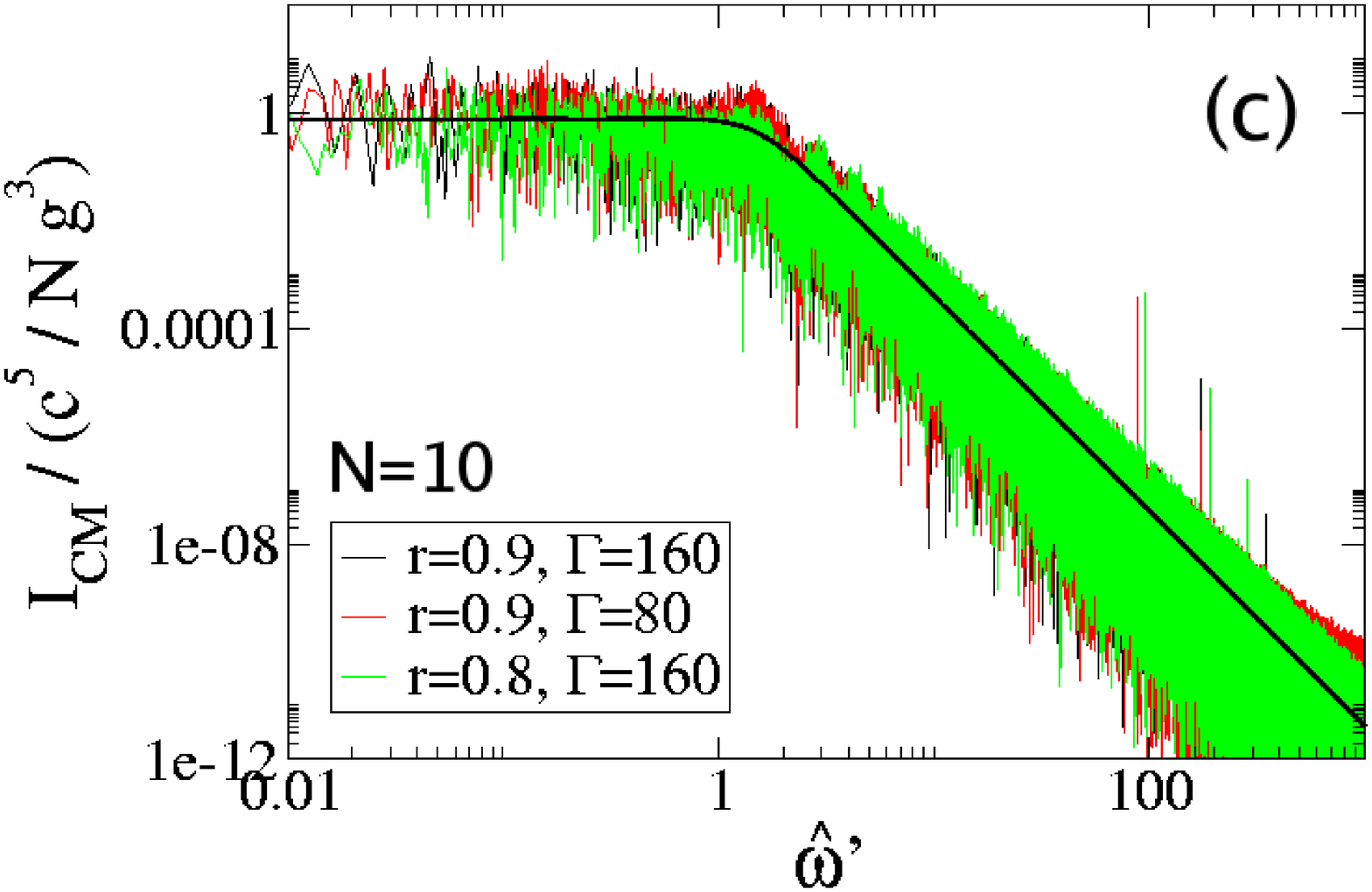}
\end{center}
\caption{\label{fig:pscom-sc} (Color online) Comparison of the power spectrum $I_{CM}$ for the height of the center of mass obtained from simulations and the proposed theory for different
$N$ values: (a) $N=1000$, (b) $N=100$, (c) $N=10$.
The thick solid line depicts the theoretical predictions given by the second term in eq.~(\ref{pscmhsc}) with $\hat{\mu}=2.0$ and $\hat{\Omega}=1.5$.}
\end{figure}  

\noindent
where $\hat{\omega}'$ is the angular frequency $\omega'$ scaled by $g/c$: $\hat{\omega}'=\omega'c/g$.
The first term in eq.~(\ref{pscmhsc}) gives the delta functional peaks at $\hat{\omega}'=\pm \hat{\omega}$ while the second term gives a continuous spectrum; these characteristics 
of the power spectrum have already been observed in refs.~\cite{clement93,luding94,warr95}.
In Fig.~\ref{fig:pscom-sc}, the scaled power spectrum successfully collapsed onto a single master curve that agreed well with the curve given by the second term in eq.~(\ref{pscmhsc}) having the curve-fit numerical factors $\hat{\mu}=2.0$ and $\hat{\Omega}=1.5$. We stress here 
that the curve given by
the second term in eq.~(\ref{pscmhsc}) is independent of the system parameters 
$N$, $r$, and $\Gamma$.

\section{Conclusions}
This paper has studied a fluidized state of a one-dimensional vibrated granular media,
using a Langevin approach. We derived a Langevin equation of motion of the COM assuming 
the forces acting on the granular fluid due to the bottom plate
under phenomenological consideration.
Using the solution of the Langevin equation, we were able to obtain
analytical expressions for several quantities, the amplitude
of the motion of the COM, the power input from the bottom plate, and
the power spectrum for the height of the COM.

In our theory, properties of the granular fluid are characterized only by
two macroscopic quantities, the total mass $M$ and the thermal velocity $c$.
Because inelastic collisions between particles are not explicitly handled
in our theory, the restitution coefficient $r$ and the number of dissipative
contacts $N-1$ in the system do not appear in the Langevin equation (\ref{lang2}).
These parameters control the thermal velocity $c$ and influence indirectly
other macroscopic quantities. 
In the previous theoretical studies~\cite{warr95,kumaran98,kumaran98-2},
the scaling relationship of the granular temperature $T$,
which is related to the thermal velocity $c$ by $c=\sqrt{k_B T/m}$,
on the system parameters including $r$ and $N$ has been obtained 
in the following way:
They estimated on the basis of kinetic theory 
the power input from the bottom plate and the rate of 
the energy dissipation by inelastic collisions between particles. 
A balance between these two quantities at steady state determines
a scaling of $T$ on the system parameters.
Since our theory can not estimate the energy dissipation by particle-particle collisions,
it is incapable of deriving the dependence of $c$ on $r$ and $N$. 

We assumed that $\left(\tau/\tau_{rel}\right)^2=\left(g/\omega c\right)^2 \ll 1$, which assured that the system was in a fluidized state, and $A_0\omega/c \ll 1$, which allowed us to use hydrodynamic sound-wave theory.
Since the acceleration $\Gamma=A_0\omega^2/g>1$ in the fluidized state, 
we have $g/\omega c<A_0\omega/c$, which implies that the second condition is dominant.

We performed simulations with $\Gamma\ge 10$ and found that if $A_0\omega/c\alt 1$, 
which corresponds to the case $\sqrt{X}\alt 1$, 
the results agreed well with the theoretical predictions.
They support our phenomenological argument deriving 
the Langevin equation (\ref{lang2}), and
our assumption that the granular fluid is 
well characterized by the total mass and the thermal
velocity within the range of validity of our theory.
\\

\noindent
{\bf Acknowledgment}

We are grateful to Professor H. Nakanishi, Dr. N. Mitarai and Dr. R. Kawahara for valuable discussions.
This study was supported by Grant-in-Aid for Scientific Research from the Ministry of Education, Culture, Sports, Science and Technology No. 19740236.
Part of the computations for this study was performed using the facilities of the Supercomputer Center, Institute for Solid State Physics, the University of Tokyo.


\begin{thebibliography}{99}

\bibitem{aranson_2006}
I. S. Aranson and L. S. Tsimring:
Rev. Mod. Phys. \textbf{78} (2006) 641.

\bibitem{clement93}
E. Cl\'{e}ment, S. Luding, A. Blumen, J. Rajchenbach, and J. Duran:
Int. J. Mod. Phys. B \textbf{7} (1993) 1807.

\bibitem{luding94}
S. Luding, E. Cl\'{e}ment, A. Blumen, J. Rajchenbach, and J. Duran:
Phys. Rev. E \textbf{49} (1994) 1634.

\bibitem{goldshtein95}
A. Goldshtein, M. Shapiro, L. Moldavsky, and M. Fichman:
J. Fluid Mech. \textbf{287} (1995) 349.

\bibitem{esipov99}
C. Salue\~{n}a, T. P\"{o}schel, and S. E. Esipov:
Phys. Rev. E \textbf{59} (1999) 4422.

\bibitem{warr95}
S. Warr, J. M. Huntley, and G. T. H. Jacques:
Phys. Rev. E \textbf{52} (1995) 5583.

\bibitem{wildman01}
R. D. Wildman, J. M. Huntley, and D. J. Parker:
Phys. Rev. E \textbf{63} (2001) 061311.

\bibitem{luding94-2}
S. Luding, H. J. Herrmann, and A. Blumen:
Phys. Rev. E \textbf{50} (1994) 3100.

\bibitem{luding95}
S.\ Luding:
Phys. Rev. E \textbf{52} (1995) 4442.

\bibitem{mcnamara98}
S. McNamara and S. Luding:
Phys. Rev. E \textbf{58} (1998) 813.

\bibitem{bernu94}
B. Bernu, F. Delyon, and R. Mazighi:
Phys. Rev. E \textbf{50} (1994) 4551.

\bibitem{kumaran98}
V.\ Kumaran:
Phys. Rev. E \textbf{57} (1998) 5660.

\bibitem{kumaran98-2}
V. Kumaran:
J. Fluid Mech. \textbf{364} (1998) 163.

\bibitem{lee95}
J. Lee:
Physica A \textbf{219} (1995) 305.

\bibitem{lee97}
J. Lee:
Physica A \textbf{238} (1997) 129.

\bibitem{bernu90}
B. Bernu and R. Mazighi:
J. Phys. A \textbf{23} (1990) 5745.

\bibitem{kubo85}
R. Kubo, M. Toda, and N. Hashitsume:
{\it Statistical Physics II} (Springer, Berlin, 1985).

\bibitem{comment1}
This analogy may seem to be unclear at first sight because the bottom plate 
in our system is assumed to oscillate sinusoidally and 
does not exhibit Brownian motion as a Brownian particle does in a fluid.
This assumption on the motion of the bottom plate, however, is idealization 
that the motion of the bottom plate is not influenced by the 
force exerted by the granular fluid. This idealization can be realized 
when we drive the bottom plate of mass $m_0$ by a sinusoidally oscillating 
external force with amplitude proportional to $m_0$, and take a limit $m_0 \to \infty$.
In the theory of Brownian motion based on the Langevin equation~\cite{kubo85}, 
forces acting on a Brownian 
particle, such as a random force and a frictional force given by Stokes' formula, 
are independent of the mass of the Brownian paritcle.
Therefore, we considered the analogy was valid even when the limit 
$m_0 \to \infty$ was taken.

\bibitem{landau87}
L. D. Landau and E. M. Lifshitz:
{\it Fluid Mechanics} (Pergamon Press, New York, 1987).

\bibitem{mcnamara92}
S. McNamara and W. R. Young:
Phys. Fluids A \textbf{4} (1992) 496.

\bibitem{mcnamara97}
S. McNamara and J. -L. Barrat:
Phys. Rev. E \textbf{55} (1997) 7767.

\bibitem{soto04}
R. Soto:
Phys. Rev. E \textbf{69} (2004) 061305.


\end{thebibliography}
\end{document}